\documentclass[12pt]{article}

\usepackage{amsmath}
\usepackage{graphics}
\usepackage{epsfig}
\usepackage{cite}
\usepackage{colordvi}
\usepackage{color}
\input epsf

\titlepage
\author{S.P.~Baranov$^1$, A.V.~Lipatov$^{2,\,3}$}
\title{$\chi_{c1}$ and $\chi_{c2}$ polarization as a probe of color octet channel}

\begin{document}

\maketitle

\begin{center}
{\it $^1$P.N.~Lebedev Institute of Physics, Moscow 119991, Russia}\\
{\it $^2$Skobeltsyn Institute of Nuclear Physics, Lomonosov Moscow State University, Moscow 119991, Russia}\\
{\it $^3$Joint Institute for Nuclear Research, Dubna 141980, Moscow Region, Russia}\\

\end{center}

\vspace{0.5cm}

\begin{center}

{\bf Abstract }

\end{center}

We analyze the first LHC data on $\chi_{c1}$ and $\chi_{c2}$ polarization 
obtained very recently by the CMS Collaboration at $\sqrt s = 8$~TeV.
We describe the perturbative production of $c\bar c$ pair
with $k_T$-factorization approach and use
nonrelativistic QCD formalism for the formation of bound states.
We demonstrate that the polar anisotropy of $\chi_{c1}$ and $\chi_{c2}$ mesons
is strongly sensitive to the color octet contributions.
We extract the long-distance matrix elements for $\chi_{c1}$ and $\chi_{c2}$ mesons
from the first CMS polarization measurement 
together with available LHC data on the $\chi_{c1}$ and $\chi_{c2}$
transverse momentum distributions (and their ratios) collected at $\sqrt s = 7$~TeV.
Our fit points to unequal color singlet wave functions of $\chi_{c1}$ and $\chi_{c2}$ 
states.

\vspace{1.0cm}

\noindent
PACS number(s): 12.38.-t, 13.20.Gd, 14.40.Pq

\newpage

Very recently, the CMS Collaboration reported on the first measurement\cite{1} of the
polarization of prompt $\chi_{c1}$ and $\chi_{c2}$ mesons produced
in $pp$ collisions at the energy $\sqrt s = 8$~TeV.
The polarizations were measured in the decay $J/\psi$ helicity frame
through the analysis of the $\chi_{c2}$ to $\chi_{c1}$ yield ratio 
as a function of the positive muon polar or azimuthal angle
in the cascade $\chi_{cJ} \to J/\psi (\to \mu^+\mu^-) + \gamma$
in three bins of $J/\psi$ transverse momentum.
No difference has been seen between the $\chi_{c1}$ and $\chi_{c2}$ states
in the azimuthal distributions,
whereas they were observed to have significantly different polar anisotropies.
Thus, at least one of these mesons should be strongly polarized along the 
helicity axis\cite{1}.
This result contrasts with the unpolarized scenario observed for direct $S$-wave 
charmonia ($J/\psi$, $\psi^\prime$) and bottomonia $\Upsilon(nS)$ at the LHC 
over a wide transverse momentum range (see, for example,\cite{2,3} and references therein).

A commonly accepted framework for the description of heavy quarkonia production and decay
is the non-relativistic Quantum Chromodynamic (NRQCD)\cite{4,5}.
The perturbatively calculated cross sections for the short distance 
production of a heavy quark pair $Q\bar Q$ in an intermediate state $^{2S + 1}L_J^{(a)}$ 
with spin $S$, orbital angular momentum $L$, total angular momentum $J$, and color 
representation $a$ are accompanied with long distance matrix elements (LDMEs) which
describe the non-perturbative transition of intermediate $Q\bar Q$ pair into a physical meson
via soft gluon radiation. The NRQCD calculations at next-to-leading order (NLO) successfully describe
charmonia $J/\psi$, $\psi^\prime$, $\chi_{cJ}$\cite{6,7,8,9,10,11,12,13} and bottomonia $\Upsilon(nS)$, 
$\chi_{bJ}(mP)$\cite{14,15,16,17,18} transverse momenta distributions
and agree well with the first CMS data\cite{1} on the $\chi_{cJ}$ polarization at the LHC.
However, NRQCD has a long-standing challenge in the $S$-wave charmonia polarization 
(see, for example, discussions\cite{19,20,21} and references therein). 
The description of $\eta_c$ production data\cite{22} reported recently 
by the LHCb Collaboration also turned out to be rather puzzling\cite{23,24}. 
So, at present the overall situation is still far from through understanding, and
further theoretical studies are still an urgent task.

One possible solution 
has been proposed in \cite{25}.
This solution implies certain modification 
of the NRQCD rules.
Usually, the final state gluons changing the color and other quantum 
numbers of quark pair and bringing it to the observed color singlet (CS)
state are regarded as carrying no energy-momentum. This is in obvious contradiction
with confinement which prohibits the emission of infinitely soft colored quanta.
In reality, the heavy quark system must undergo a kind of final state interaction 
where the energy-momentum exchange must be larger than at least the typical 
confinement scale. Then, the classical multipole radiation theory can be 
applied to describe nonperturbative transformations of the color octet (CO) quark 
pairs produced in hard subprocesses into observed final state quarkonia.
In this way, the polarization puzzle for $S$-wave charmonia\cite{26} and
bottomonia\cite{27,28} and the production puzzle for $\eta_c$ mesons\cite{29} have been 
successfully solved.
Further on, a good description of the $\chi_{c1}$ and $\chi_{c2}$ production
cross sections including their relative  rates $\sigma(\chi_{c2})/\sigma(\chi_{c1})$ 
has been achieved and the corresponding LDMEs for $\chi_{cJ}$ mesons have been determined\cite{26}.

The main goal of our present note is to 
extend the approach\cite{25}
to the first and very new CMS data\cite{1} on $\chi_{cJ}$ polarization.
We propose a method to implement these data into the LDMEs fit procedure, thus  
refining the previously extracted LDMEs for $\chi_{cJ}$ mesons.
Our study sheds light on the role of CO contributions
which were unnecessary or even unwanted\cite{12} for $\chi_{cJ}$ $p_T$ spectra
or their relative rates $\sigma(\chi_{c2})/\sigma(\chi_{c1})$,
but which reveal now in the measured polar anisotropies.
To preserve the consistency with our previous studies\cite{26,27,28,29},
we follow mostly the same steps and
employ the $k_T$-factorization QCD approach\cite{30,31} 
to produce the $c\bar c$ pair in the hard parton scattering.
The newly added calculations are only for the feeddown contributions
from $\psi^\prime$ radiative decays.

For the reader's convenience, 
we briefly recall the calculation details.
Our consideration is based on the off-shell gluon-gluon fusion
subprocess that represents the
true leading order (LO) in QCD:
\begin{equation}
  g^*(k_1) + g^*(k_2) \to c\bar c\left[ ^3P_J^{[1]}, \, ^3S_1^{[8]} \right](p)
\end{equation}
\noindent
for $\chi_{cJ}$ mesons with $J = 0$, $1$, $2$. 
The four-momenta of all particles are indicated in the parentheses and 
the possible intermediate states of the $c\bar c$ pair are listed in the brackets.
The initial off-shell gluons have non-zero transverse momenta
$k_{1T}^2 = - {\mathbf k}_{1T}^2 \neq 0$,
$k_{2T}^2 = - {\mathbf k}_{2T}^2 \neq 0$
and, consequently, an admixture of longitudinal component in the polarization vectors.
According to the $k_T$-factorization prescription\cite{31}, 
the gluon spin density matrix is taken in the form
\begin{equation}
  \sum \epsilon^\mu \epsilon^{*\nu} = { {\mathbf k}_{T}^\mu {\mathbf k}_{T}^\nu \over {\mathbf k}_{T}^2},
\end{equation}
\noindent
where ${\mathbf k}_T$ is the component of the gluon momentum perpendicular to the beam axis.
In the collinear limit, where ${\mathbf k}_T^2 \to 0$, this expression converges to the ordinary 
$-g^{\mu\nu}$
after averaging over the gluon azimuthal angle.
In all other respects, we follow the standard QCD Feynman rules. The hard
production amplitudes contain spin and color projection operators\cite{32} 
that guarantee the proper quantum numbers of the state under consideration.
The respective cross section
\begin{equation}
  \displaystyle \sigma(pp \to \chi_{cJ} + X) = \int {2\pi \over x_1 x_2 s F} \, f_g(x_1, {\mathbf k}_{1T}^2, \mu^2) f_g(x_2, {\mathbf k}_{2T}^2), \mu^2) \, \times \atop {
  \displaystyle \times \, |{\cal \bar A}(g^* + g^* \to \chi_{cJ})|^2 d{\mathbf k}_{1T}^2 d{\mathbf k}_{2T}^2 dy {d\phi_1 \over 2\pi} {d\phi_2 \over 2\pi} },
\end{equation}
\noindent 
where $\phi_1$ and $\phi_2$ are the azimuthal angles of  
incoming off-shell gluons carrying 
the longitudinal momentum fractions $x_1$ and $x_2$, $y$ is the 
rapidity of produced $\chi_{cJ}$ mesons, $F$ is the off-shell flux factor\cite{34}
and $f_g(x,{\mathbf k}_{T}^2, \mu^2)$ 
is the transverse momentum dependent (TMD, or unintegrated) gluon density function.
More details can be found in our previous papers\cite{26,27,28,29}.
Presently, all of the above formalism is implemented into the newly developed
Monte-Carlo event generator \textsc{pegasus}\cite{33}.

As usual, we have tried several sets of TMD gluon densities in a proton. 
Three of them, namely, A0\cite{35}, JH'2013 set 1 and JH'2013 set 2\cite{36} have been obtained
from Catani-Ciafaloni-Fiorani-Marchesini (CCFM) evolution equation\cite{37}, 
where the input parametrizations (used as boundary conditions) have been 
fitted to the proton structure function $F_2(x,Q^2)$. 
Besides that, we have tested a TMD gluon distribution obtained within the 
Kimber-Martin-Ryskin (KMR) prescription\cite{38,39},
which provides a method to construct the TMD parton 
densities from conventional (collinear) ones\footnote{For the input, we have used LO MMHT'2014 set\cite{40}.}. 
Following\cite{41}, we set the meson masses to
$m(\chi_{c1}) = 3.51$~GeV, $m(\chi_{c2}) = 3.56$~GeV,
$m(J/\psi) = 3.096$~GeV and branching fractions
$B(\chi_{c1} \to J/\psi \gamma) = 33.9$\%, 
$B(\chi_{c2} \to J/\psi \gamma) = 19.2$\% and
$B(J/\psi \to \mu^+\mu^-) = 5.961$\% everywhere in the calculations below.
When evaluating the feeddown 
contributions from $\psi^\prime$ radiative decays, $\psi^\prime \to \chi_{cJ} + \gamma$,
we set $m(\psi^\prime) = 3.69$~GeV,
$B(\psi^\prime \to \chi_{c1} \gamma) = 9.75$\% and
$B(\psi^\prime \to \chi_{c2} \gamma) = 9.52$\%.
The parton level calculations 
have been performed using the Monte-Carlo generator \textsc{pegasus}. 

As it was mentioned above, to determine the LDMEs of $\chi_{cJ}$ mesons 
a global fit to the $\chi_{cJ}$ production data at the LHC was performed\cite{26}. 
The data on the $\chi_{c1}$ and $\chi_{c2}$ transverse momentum distributions provided by 
ATLAS Collaboration\cite{42} at $\sqrt s = 7$~TeV and the production rates 
$\sigma(\chi_{c2})/\sigma(\chi_{c1})$ reported by CMS\cite{43}, ATLAS\cite{42} and 
LHCb\cite{44,45} Collaborations were included in the fit.
Here we extend our previous consideration and incorporate it with
the first data\cite{1} on the $\chi_{c1}$ and $\chi_{c2}$
polarization collected by CMS Collaboration at $\sqrt s = 8$~TeV.
In the original CMS analysis, the $\chi_{cJ}$ polarization was extracted from the (di)muon 
angular distributions in the helicity frame of the daughter $J/\psi$ meson.
The latter is parametrized as
\begin{equation}
 {d\sigma\over d\cos\theta^*\,d\phi^*}\sim{1\over 3+\lambda_\theta}
 \left(1+\lambda_\theta\cos^2\theta^*+\lambda_\phi\sin^2\theta^*\cos 2\phi^*
 +\lambda_{\theta\phi}\sin2\theta^*\cos\phi^* \right),
\end{equation}
\noindent
where $\theta^*$ and $\phi^*$ are the positive muon polar and azimuthal angles,
so that the $\chi_{cJ}$ angular momentum is encoded 
in the polarization parameters $\lambda_\theta$, $\lambda_\phi$ and $\lambda_{\theta \phi}$.
The ratio of the yields $\sigma(\chi_{c2})/\sigma(\chi_{c1})$
has been measured as a function of $\cos\theta^*$ and $\phi^*$ in three different regions
of $J/\psi$ transverse momentum, $8 < p_T < 12$~GeV, $12 < p_T < 18$~GeV and $18 < p_T < 30$~GeV,
thus leading to a simple correlation between the 
$\lambda_{\theta}^{\chi_{c1}}$ and $\lambda_{\theta}^{\chi_{c2}}$
parameters:
\begin{equation}
  \lambda_{\theta}^{\chi_{c2}} = \left( - 0.94 + 0.90\lambda_{\theta}^{\chi_{c1}}\right) \pm \left(0.51 + 0.05\lambda_{\theta}^{\chi_{c1}}\right), \quad 8 < p_T < 12~{\rm GeV},
\end{equation}
\begin{equation}
  \lambda_{\theta}^{\chi_{c2}} = \left( - 0.76 + 0.80\lambda_{\theta}^{\chi_{c1}}\right) \pm \left(0.26 + 0.05\lambda_{\theta}^{\chi_{c1}}\right), \quad 12 < p_T < 18~{\rm GeV},
\end{equation}
\begin{equation}
  \lambda_{\theta}^{\chi_{c2}} = \left( - 0.78 + 0.77\lambda_{\theta}^{\chi_{c1}}\right) \pm \left(0.26 + 0.06\lambda_{\theta}^{\chi_{c1}}\right), \quad 18 < p_T < 30~{\rm GeV}.
\end{equation}
\noindent 
Our main idea is to extract the LDME for $^3S_1^{[8]}$ contributions, ${\cal O}^{\chi_{c0}} \big[ \, ^3S_1^{[8]} \big]$,
from the polarization data, since it can only be poorly determined from the measured
$\chi_{cJ}$ transverse momentum distributions. To be precise, a
good description of the latter can be achieved for a widely ranging
${\cal O}^{\chi_{c0}} \big[ \, ^3S_1^{[8]} \big]$, always with 
reasonably good $\chi^2/d.o.f.$ (see, for example,\cite{11,12,13}).
Moreover, its zero value is even preferable for the production rate ratio
$\sigma(\chi_{c2})/\sigma(\chi_{c1})$\cite{12}.
However, the reported production rates plotted as functions of $\cos\theta^*$ and $\phi^*$
have free (indefinite) normalization\cite{46} and thus it is difficult
to immediately implement them into the LDMEs fitting procedure.
Therefore, we had to use the parametrizations (5) --- (7) for our purposes.

Our fitting procedure is the following.
First, we performed a fit of the
$\chi_{c1}$ and $\chi_{c2}$ transverse momentum distributions
and their relative production rates $\sigma(\chi_{c2})/\sigma(\chi_{c1})$
and determined the values of CS wave functions of $\chi_{cJ}$ mesons 
at the origin, $|{\cal R}^{\prime \chi_{c1}}(0)|^2$ and $|{\cal R}^{\prime \chi_{c2}}(0)|^2$,
for a (large) number of fixed guessed
${\cal O}^{\chi_{c0}} \big[ \, ^3S_1^{[8]} \big]$ values in the range 
$10^{-4} < {\cal O}^{\chi_{c0}} \big[ \, ^3S_1^{[8]} \big] < 10^{-3}$~GeV$^3$.
At this step we employ the fitting algorithm 
implemented in the \textsc{gnuplot} package\cite{47}.
Following\cite{48}, we considered the CS
wave functions as independent (not necessarily identical) free parameters. The 
reason for such a suggestion is that treating the charmed quarks 
in the potential models as spinless particles could be an oversimplification, and 
radiative corrections to the CS wave functions could be large\cite{48} and spin dependent.
Then, we collected the simulated events in the kinematical 
region defined by the CMS measurement\cite{1}
and generated the decay muon angular
distributions according to the production and decay matrix elements.
By applying a three-parametric fit based on~(4), we 
determined the polarization parameters $\lambda_{\theta}^{\chi_{c1}}$
and $\lambda_{\theta}^{\chi_{c2}}$
as functions of ${\cal O}^{\chi_{c0}} \big[ \, ^3S_1^{[8]} \big]$ (see Fig.~1).
We find that the dependence of these parameters 
on ${\cal O}^{\chi_{c0}} \big[ \, ^3S_1^{[8]} \big]$
is essential and therefore can be used to extract the latter from the data.
One can see that $\chi_{c1}$ and $\chi_{c2}$ mesons 
have significantly different polar anisotropies,
$\lambda_{\theta}^{\chi_{c1}} > 0$ and $\lambda_{\theta}^{\chi_{c2}} < 0$,
which smoothly decrease when ${\cal O}^{\chi_{c0}} \big[ \, ^3S_1^{[8]} \big]$ 
grows\footnote{The influence of CO contributions on the $\chi_{cJ}$ polarization 
in the collinear scheme has been investigated in \cite{49}.}. 
It is important to remind that each of the considered
${\cal O}^{\chi_{c0}} \big[ \, ^3S_1^{[8]} \big]$ values provides already a good fit 
to the $p_T$ spectra: each value of ${\cal O}^{\chi_{c0}} \big[ \, ^3S_1^{[8]} \big]$
is associated with a respective set of commonly fitted color-singlet LDMEs.
Now, using the relations (5) --- (7) between
$\lambda_{\theta}^{\chi_{c1}}$ and $\lambda_{\theta}^{\chi_{c2}}$
(shown by dashed curves in Fig.~1) one can easily extract 
${\cal O}^{\chi_{c0}} \big[ \, ^3S_1^{[8]} \big]$ for 
each of the three $p_T$ regions.
Finally, the mean-square average is taken as the fitted value.
Thus, this provides us with a complementary way 
to determine the LDMEs for $\chi_{cJ}$ mesons from the polarization data.

It is interesting to note that the determined values of ${\cal O}^{\chi_{c0}} \big[ \, ^3S_1^{[8]} \big]$
almost do not depend on the exact polarization of $^3S_{1}^{[8]}$ contributions
in the CO channel.
This can be easily understood because $\chi_{c1}$ and $\chi_{c2}$ mesons 
from the $^3S_1^{[8]}$ intermediate state 
produce very close $J/\psi$ polarization,
while the measured polar asymmetry is driven by the 
difference $\lambda_{\theta}^{\chi_{c1}}-\lambda_{\theta}^{\chi_{c2}}$.
To illustrate it, we have repeated the calculations treating 
the $^3S_1^{[8]}$ contributions as unpolarized (yellow curves
in Fig.~1). As one can see, the correlations (5) --- (7) obtained in this toy approximation
practically coincide with exact calculations.

The mean-square average of the 
extracted ${\cal O}^{\chi_{c0}} \big[ \, ^3S_1^{[8]} \big]$ values and the 
corresponding CS wave functions
at the origin $|{\cal R}^{\prime \chi_{c1}}(0)|^2$ and $|{\cal R}^{\prime \chi_{c2}}(0)|^2$
are shown in Table~1 for all tested TMD gluon densities.
The relevant uncertainties are estimated in the conventional way
using Student's t-distribution at the confidence level $P = 95$\%.
For comparison, we also present the LDMEs
obtained in the NLO NRQCD by other authors\cite{12,13}.
Our fit shows unequal values 
for the $\chi_{c1}$ and $\chi_{c2}$ wave functions
with the ratio $|{\cal R}^{\prime\chi_{c1}}(0)|^2/|{\cal R}^{\prime\chi_{c2}}(0)|^2 \sim 4$
for CCFM-evolved TMD gluon densities and about of $|{\cal R}^{\prime\chi_{c1}}(0)|^2/|{\cal R}^{\prime\chi_{c2}}(0)|^2 \sim 3$ for KMR one.
Thus, we interpret the available LHC data as supporting their unequal values,
that qualitatively agrees with the previous results\cite{26,48}.
This leads to a different role of CO contributions to the $\chi_{c1}$ and $\chi_{c2}$ 
production cross sections. So, the $\chi_{c1}$ production
is dominated by the CS contributions, whereas CO terms are more important 
for $\chi_{c2}$ mesons (see Fig.~2).

\begin{table}
\caption{The fitted values of LDMEs and CS wave functions at the origin
for $\chi_{cJ}$ mesons.
The results obtained in the NLO NRQCD fits\cite{12,13} are shown for comparison.}
\begin{tabular}{lccc}
\hline \hline \\
 Source & \quad $|{\cal R}^{\prime \chi_{c1}}(0)|^2$/GeV$^5$ \quad & \quad $|{\cal R}^{\prime \chi_{c2}}(0)|^2$/GeV$^5$ \quad & \quad ${\cal O}^{\chi_{c0}} \big[ \, ^3S_1^{[8]} \big]$/GeV$^3$ \quad \\
\\ \hline \hline
\\
  A0 & $0.14 \pm 0.03$ & $0.0346 \pm 0.0010$ & $(7.0 \pm 2.0) \cdot 10^{-4}$ \\
\\  
  JH'2013 set 1 & $0.17 \pm 0.03$ & $0.043 \pm 0.004$ & $(7.0 \pm 2.0) \cdot 10^{-4}$ \\
\\
  JH'2013 set 2 & $0.20 \pm 0.04$ & $0.0500 \pm 0.0007$ & $(8.0 \pm 2.0) \cdot 10^{-4}$ \\
\\  
  KMR (MMHT'2014) & $0.08 \pm 0.02$ & $0.026 \pm 0.002$ & $(4.0 \pm 1.0) \cdot 10^{-4}$ \\
\\
  NLO NRQCD fit\cite{12} & $0.35$ & $0.35$ & $4.4 \cdot 10^{-4}$ \\
\\  
  NLO NRQCD fit\cite{13} & $0.075$ & $0.075$ & $2.01 \cdot 10^{-3}$ \\
\\  
\hline \hline
\end{tabular}
\end{table}

All the LHC data involved in the fits are compared with our predictions in Figs.~2 --- 4.
The green shaded bands represent the theoretical uncertainties of our calculations 
(responding to JH'2013 set 2 gluon density), 
which include both the scale uncertainties and the ones coming from the 
LDMEs fitting procedure.
To estimate the scale uncertainties, the standard variations in the scale (by a factor of 2)
were applied through replacing the JH'2013 set 2 gluon density
with JH'2013 set 2$+$, or with JH'2013 set 2$-$, respectively. This was done to preserve
the intrinsic correspondence between the TMD set and scale used in the evolution
equation (see\cite{36} for more information).
We have achieved quite a nice agreement between our calculations and available 
LHC data. In particular, we obtained a simultaneous description of the 
transverse momentum distributions and the relative production rates 
$\sigma(\chi_{c2})/\sigma(\chi_{c1})$.
There are some deviations from the data at low $p_T$ region,
where, however, an accurate treatment of 
large logarithms $\ln m(\chi_{cJ})/p_T$ 
and other nonperturbative effects is needed.

The $\lambda_{\theta}^{\chi_{c2}}$ values extracted according to (5) --- (7) 
when $\lambda_{\theta}^{\chi_{c1}}$ is fixed to our predictions
are shown on Fig.~5. As one can see, our fit well agrees
with the experimentally determined correlations between $\lambda_{\theta}^{\chi_{c1}}$
and $\lambda_{\theta}^{\chi_{c2}}$.
The predicted $\lambda_{\theta}^{\chi_{cJ}}$ values are  
practically independent on the TMD gluon density and are close to 
the reported NLO NRQCD results\cite{1}.

To conclude, we have considered first LHC data on $\chi_{c1}$ and $\chi_{c2}$ polarizations 
reported very recently by the CMS Collaboration at $\sqrt s = 8$~TeV.
We have demonstrated that the polar anisotropy of $\chi_{c1}$ and $\chi_{c2}$ mesons
is strongly sensitive to the color octet contributions
and proposed a method to extract the corresponding LDMEs from the polarization data.
First time with the $k_T$-factorization approach,
we have determined the color octet LDMEs and the color 
singlet wave functions at the origin 
$|{\cal R}^{\prime \chi_{c1}}(0)|^2$ and $|{\cal R}^{\prime \chi_{c2}}(0)|^2$, thus
refining our previous results 
based on the measured $\chi_{cJ}$ transverse momentum distributions only.
Our fit points to unequal color singlet wave functions of $\chi_{c1}$ and $\chi_{c2}$ states
with $|{\cal R}^{\prime\chi_{c1}}(0)|^2/|{\cal R}^{\prime\chi_{c2}}(0)|^2 \sim 3$ or $4$.
We achieved a good simultaneous description of all available data
on $\chi_{cJ}$ production at the LHC, including their 
transverse momentum distributions, relative production rates and polarization observables.

{\sl Acknowledgements.}
The authors thank H.~Jung for useful discussions on the topic. We 
are grateful to DESY Directorate for the support in the framework of
Cooperation Agreement between MSU and DESY on phenomenology of the LHC processes
and TMD parton densities.

\newpage

\begin{figure}
\begin{center}
\includegraphics[width=8.1cm]{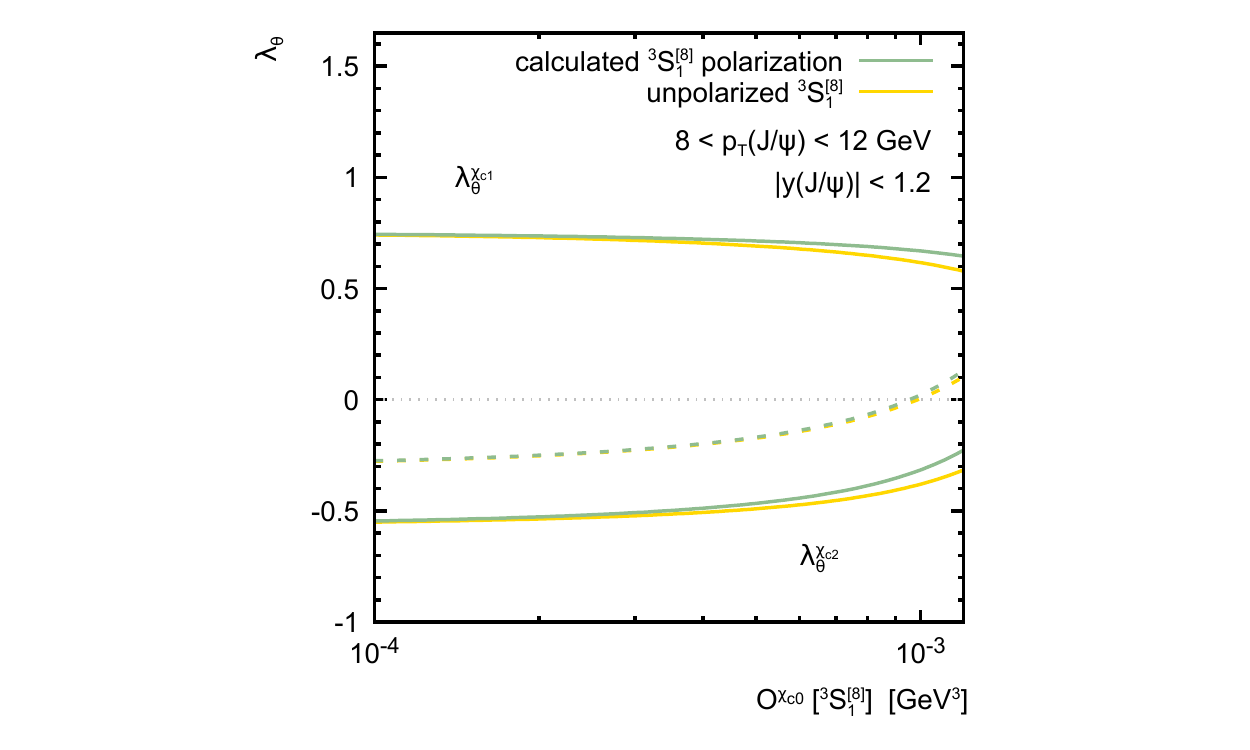}
\includegraphics[width=8.1cm]{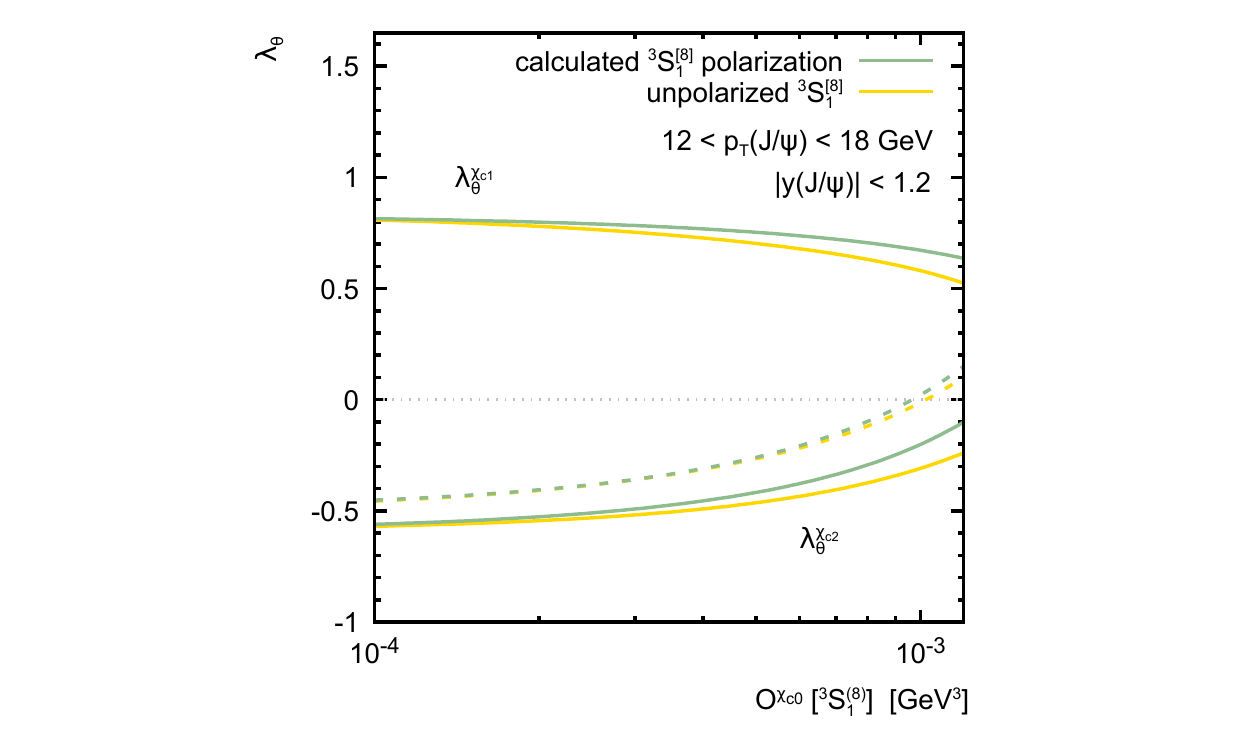}
\includegraphics[width=8.1cm]{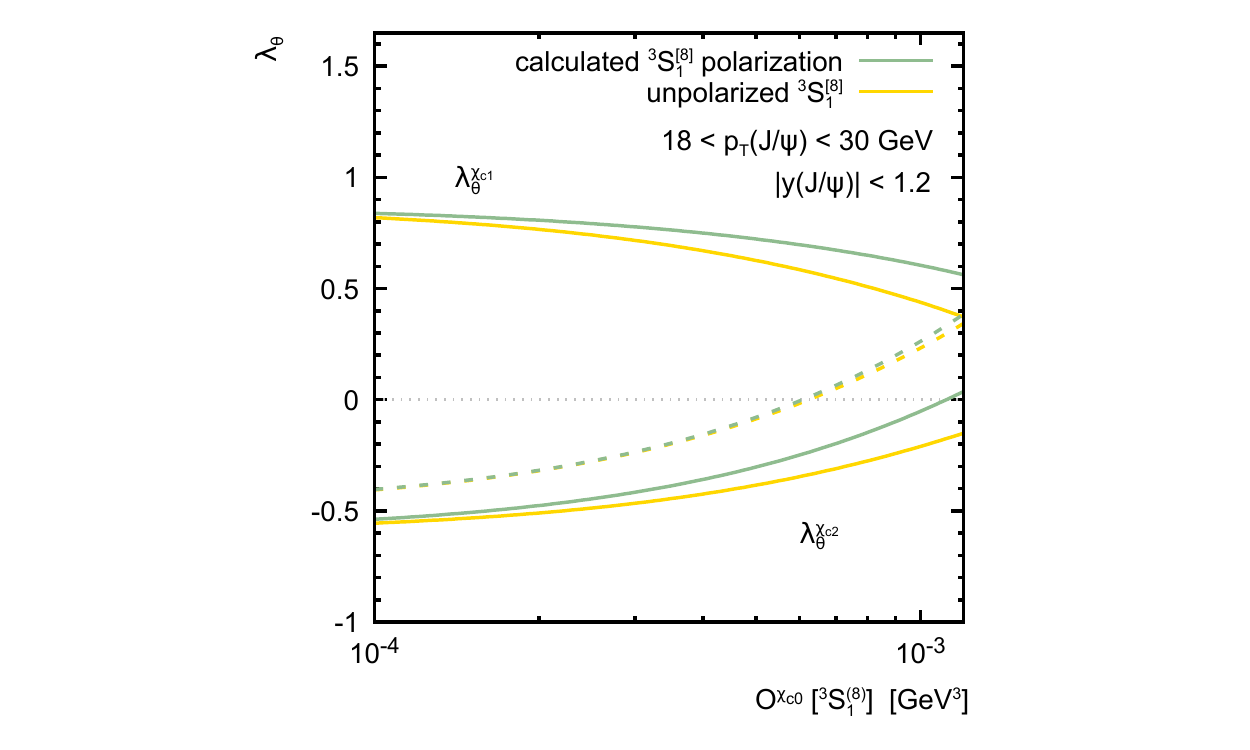}
\caption{Polarization parameters $\lambda_{\theta}^{\chi_{c1}}$
and $\lambda_{\theta}^{\chi_{c2}}$ calculated
as a functions of ${\cal O}^{\chi_{c0}} \big[ \, ^3S_1^{[8]} \big]$ in the 
helicity frame at $|y(J/\psi)| < 1.2$ and $\sqrt s = 8$~TeV in three different $p_T$ 
regions.
Solid green and yellow curves represent the results of exact and approximated 
(when the intermediate $^3S_1^{[8]}$ state is taken unpolarized) calculations. 
Dashed curves correspond to the correlations~(5) --- (7)
reported by the CMS Collaboration\cite{1}.
Everywhere, the JH'2013 set 2 gluon density is used.}
\label{fig1}
\end{center}
\end{figure}

\begin{figure}
\begin{center}
\includegraphics[width=8.1cm]{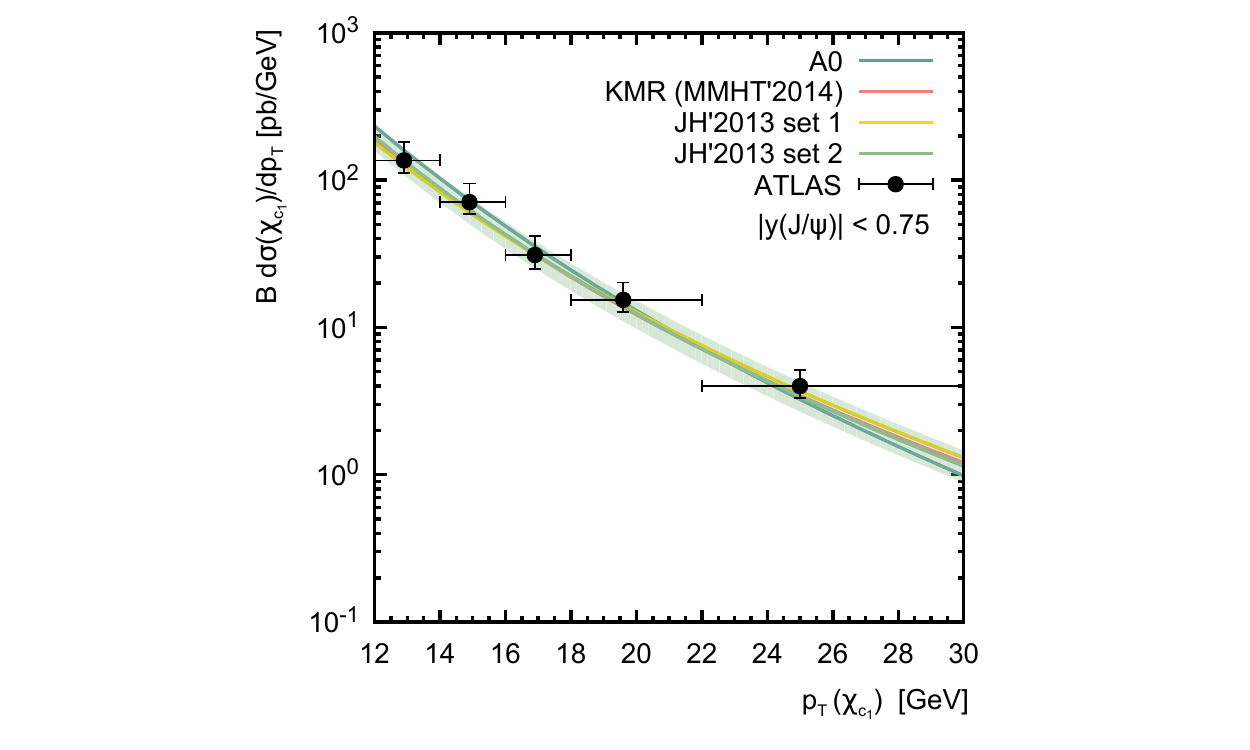}
\includegraphics[width=8.1cm]{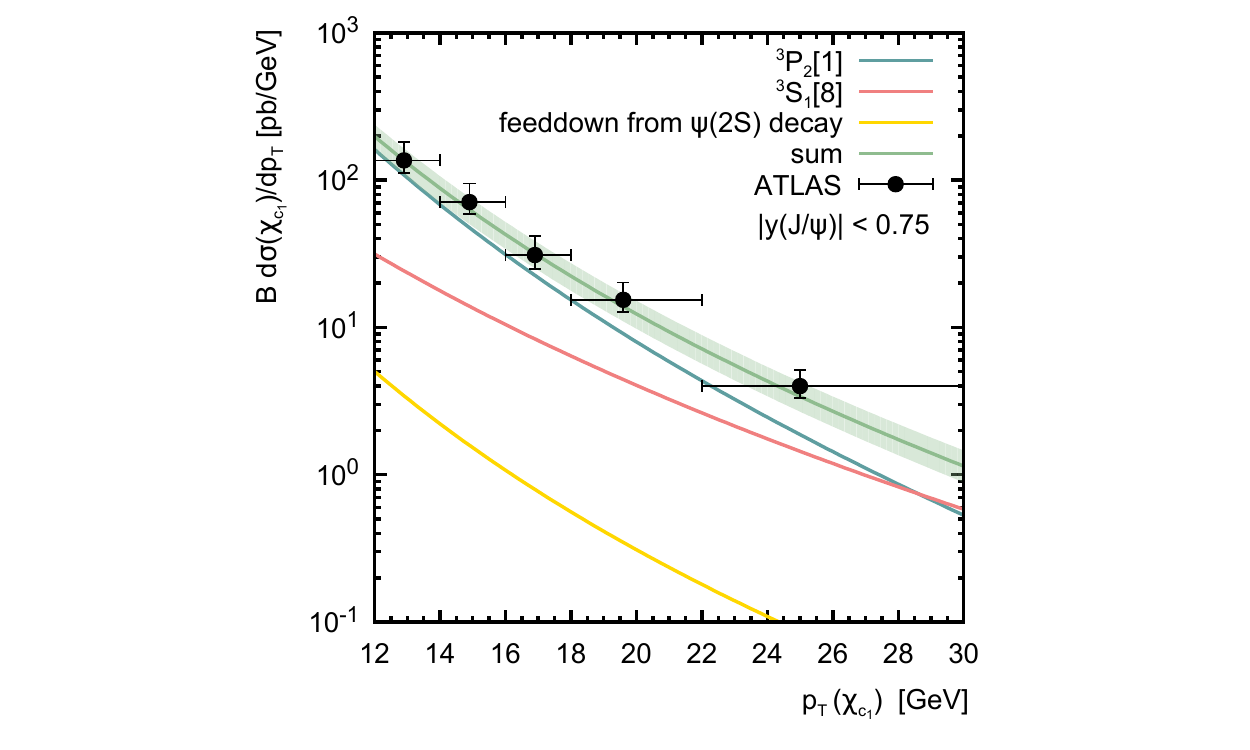}
\includegraphics[width=8.1cm]{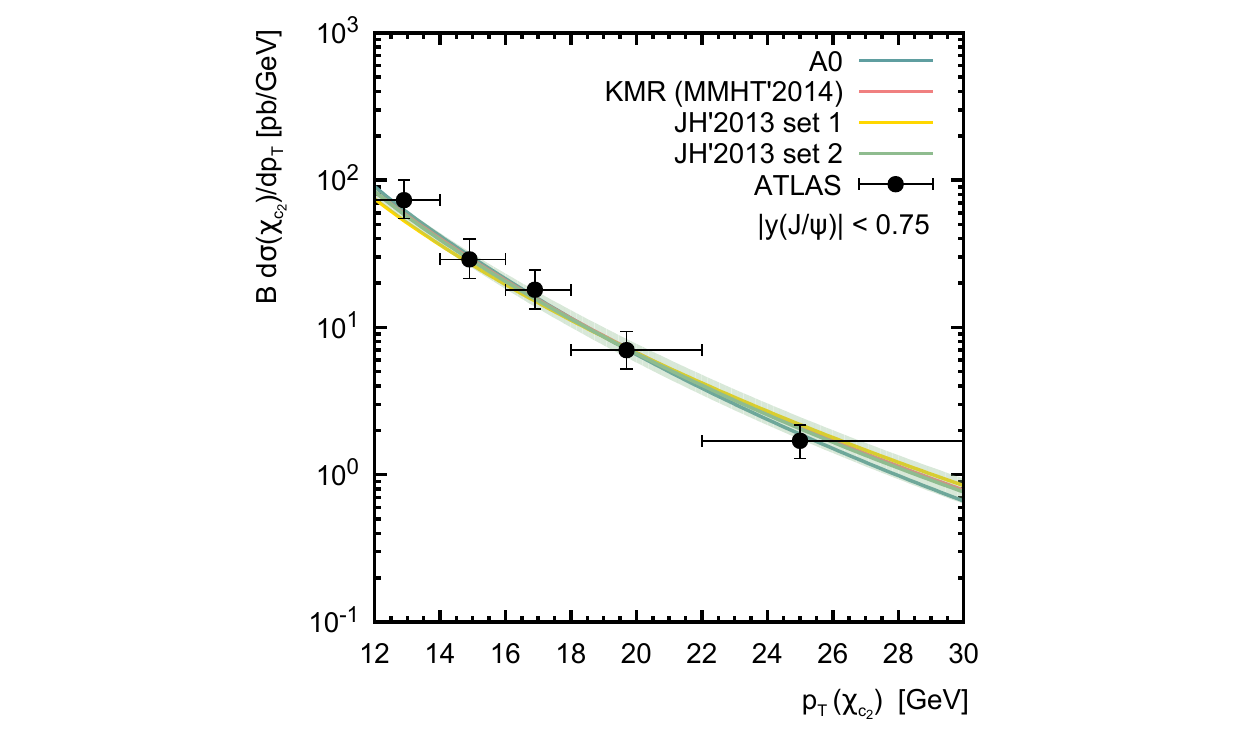}
\includegraphics[width=8.1cm]{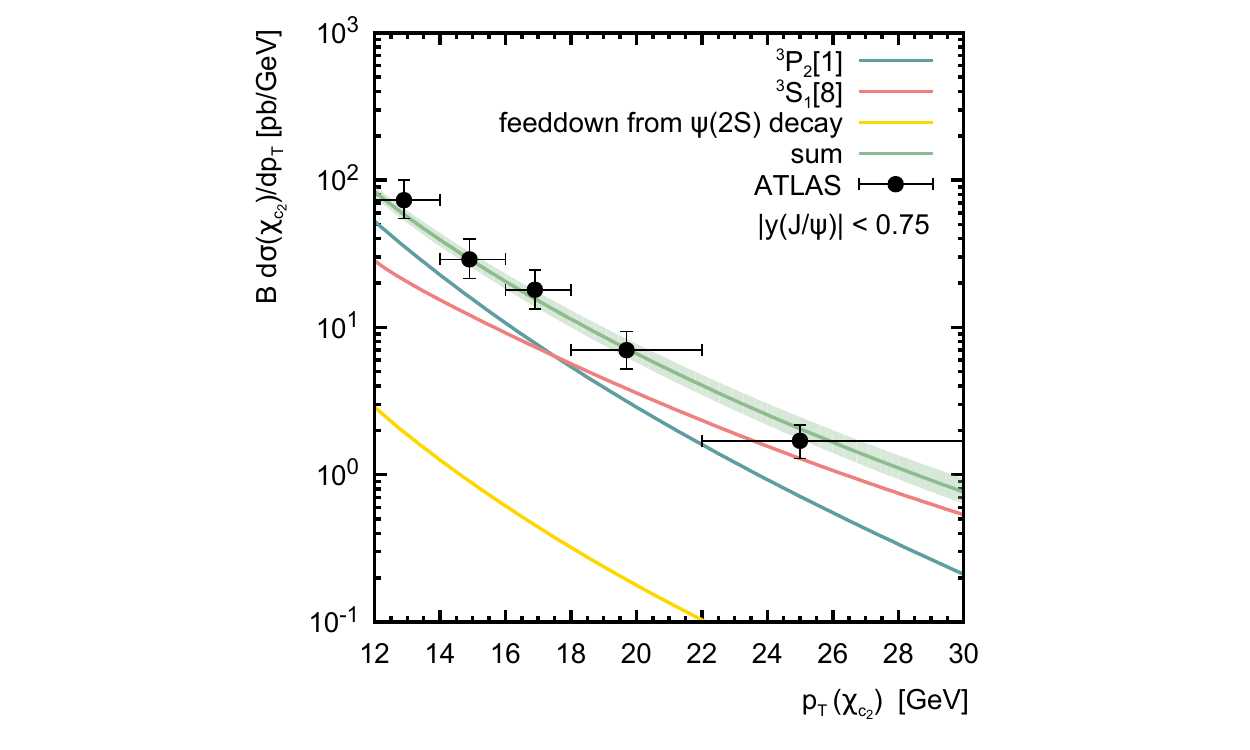}
\caption{The prompt $\chi_{c1}$ and $\chi_{c2}$ production 
cross sections in $pp$ collisions at $\sqrt s = 7$~TeV
as a function of their transverse momenta.
On left panels, the predictions obtained with
different TMD gluon densities in a proton are presented.
On right panels, the contributions from direct $^3P_{J}^{[1]}$, $^3S_1^{[8]}$ and feeddown
production mechanisms are shown separately (the JH'2013 set 2 gluon distribution
was used for illustration).
The experimental data are from ATLAS\cite{42}.}
\label{fig2}
\end{center}
\end{figure}

\begin{figure}
\begin{center}
\includegraphics[width=8.1cm]{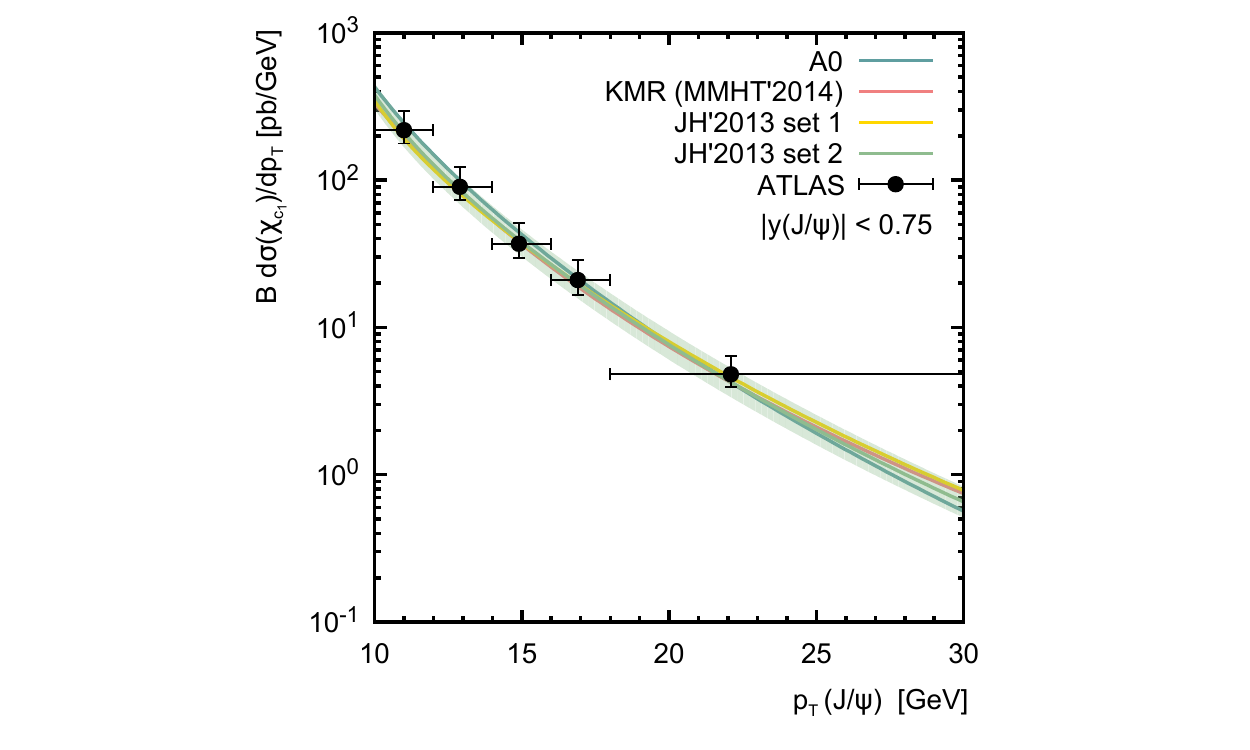}
\includegraphics[width=8.1cm]{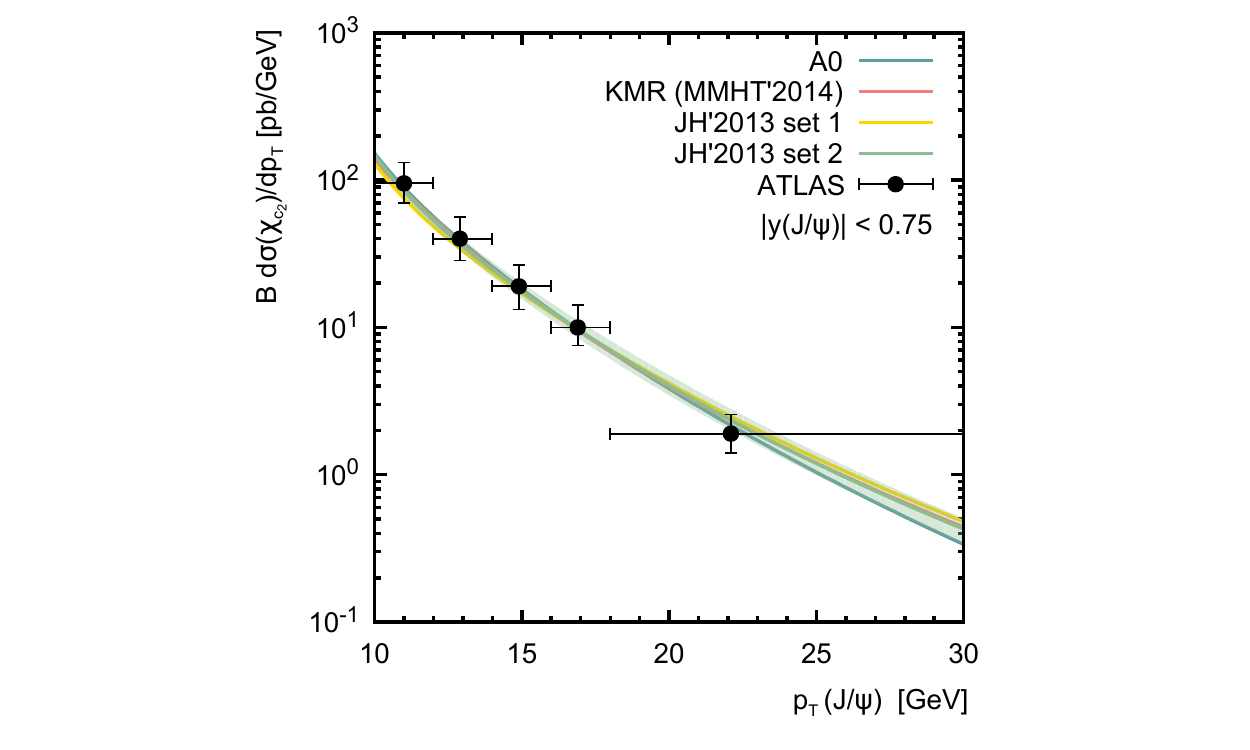}
\caption{The prompt $\chi_{c1}$ and $\chi_{c2}$ production 
cross sections in $pp$ collisions at $\sqrt s = 7$~TeV
as a function of decay $J/\psi$ transverse momenta.
Notation of all curves is the same as in Fig.~2.
The experimental data are from ATLAS\cite{42}.}
\label{fig3}
\end{center}
\end{figure}

\begin{figure}
\begin{center}
\includegraphics[width=8.1cm]{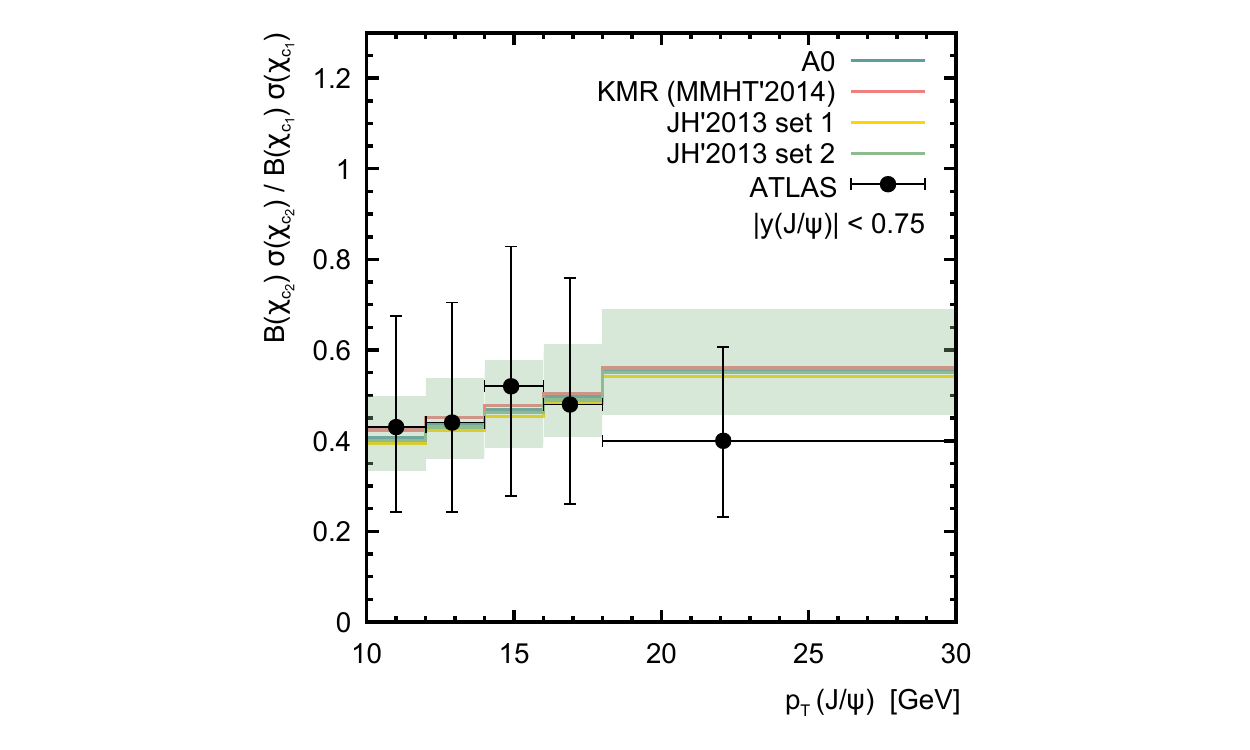}
\includegraphics[width=8.1cm]{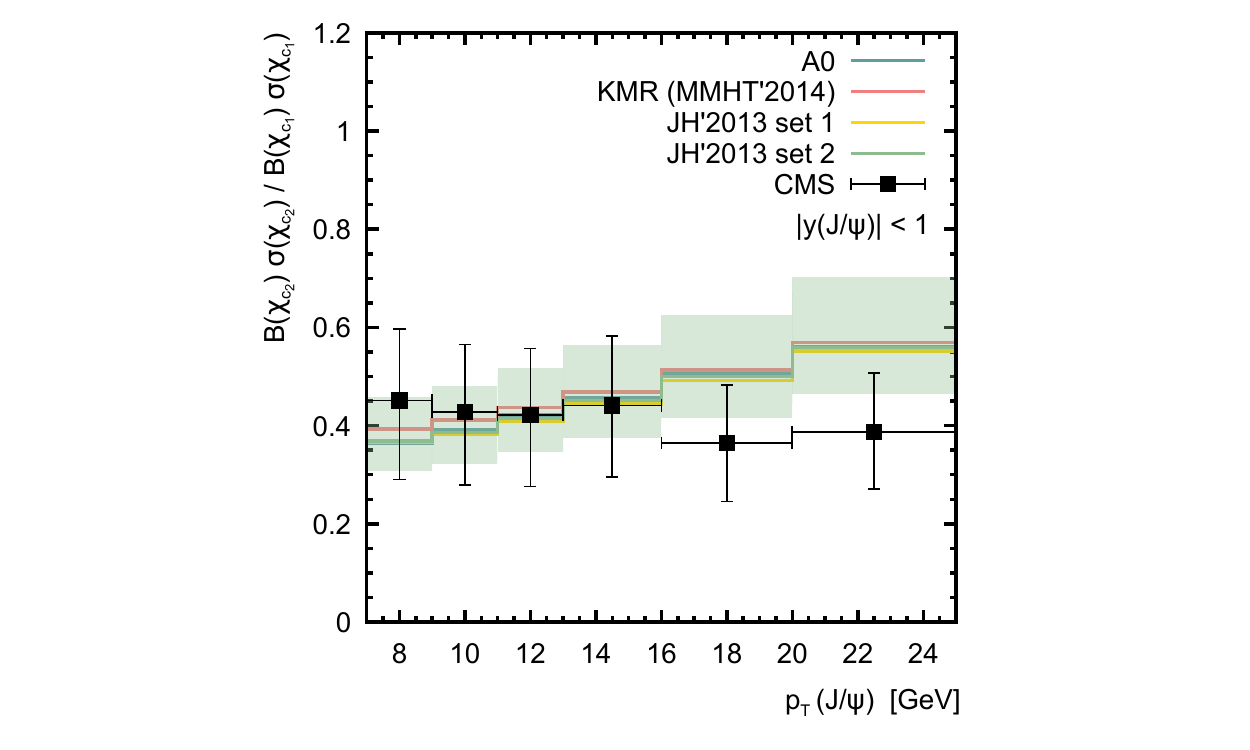} 
\includegraphics[width=8.1cm]{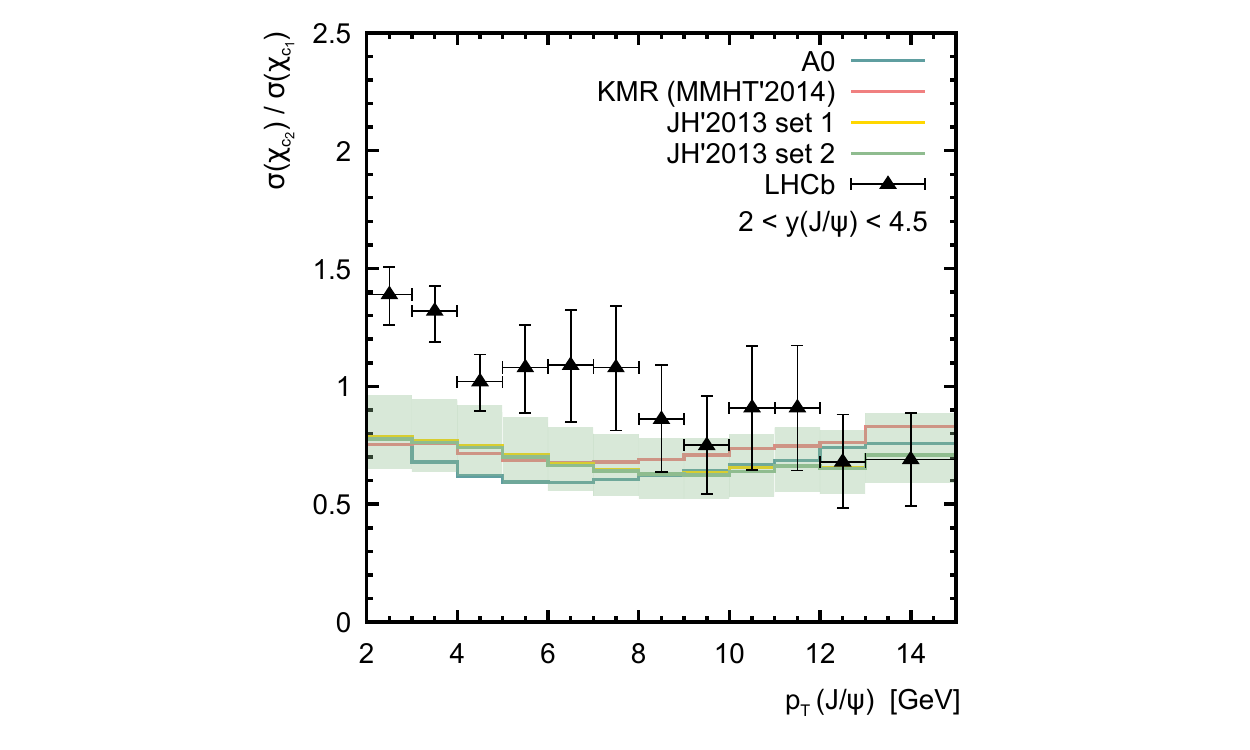} 
\includegraphics[width=8.1cm]{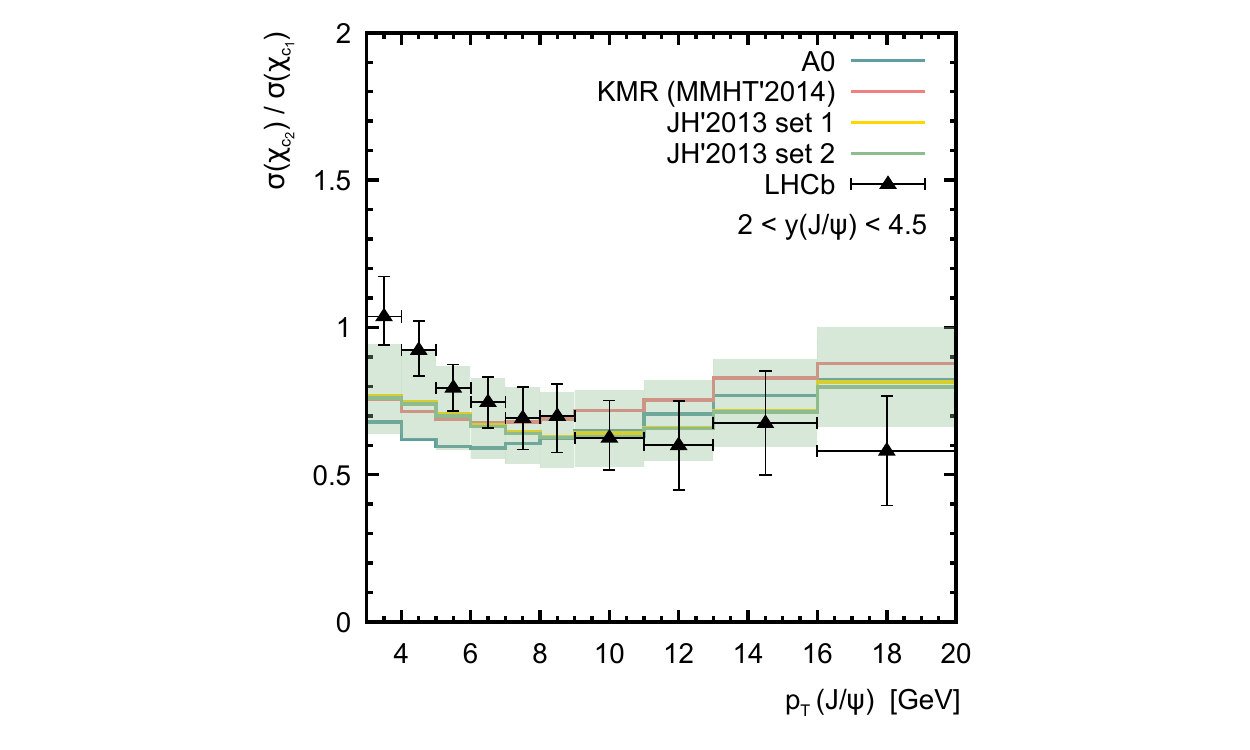} 
\caption{The relative production rate $\sigma(\chi_{c2})/\sigma(\chi_{c1})$
calculated as a function of decay $J/\psi$ transverse momentum at $\sqrt s = 7$~TeV. 
Notation of all curves is the same as in Fig.~2. The
experimental data are from ATLAS\cite{42}, CMS\cite{43} and LHCb\cite{44,45}.}
\label{fig4}
\end{center}
\end{figure}

\begin{figure}
\begin{center}
\includegraphics[width=8.1cm]{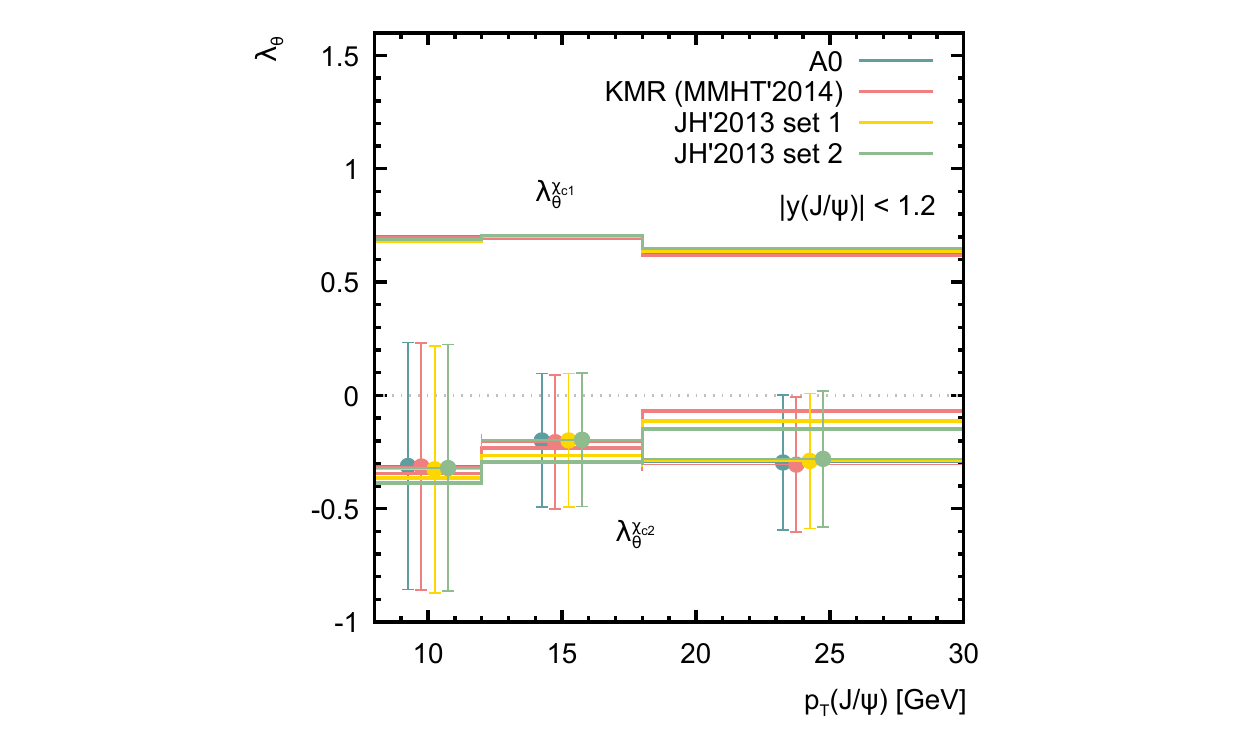}
\includegraphics[width=8.1cm]{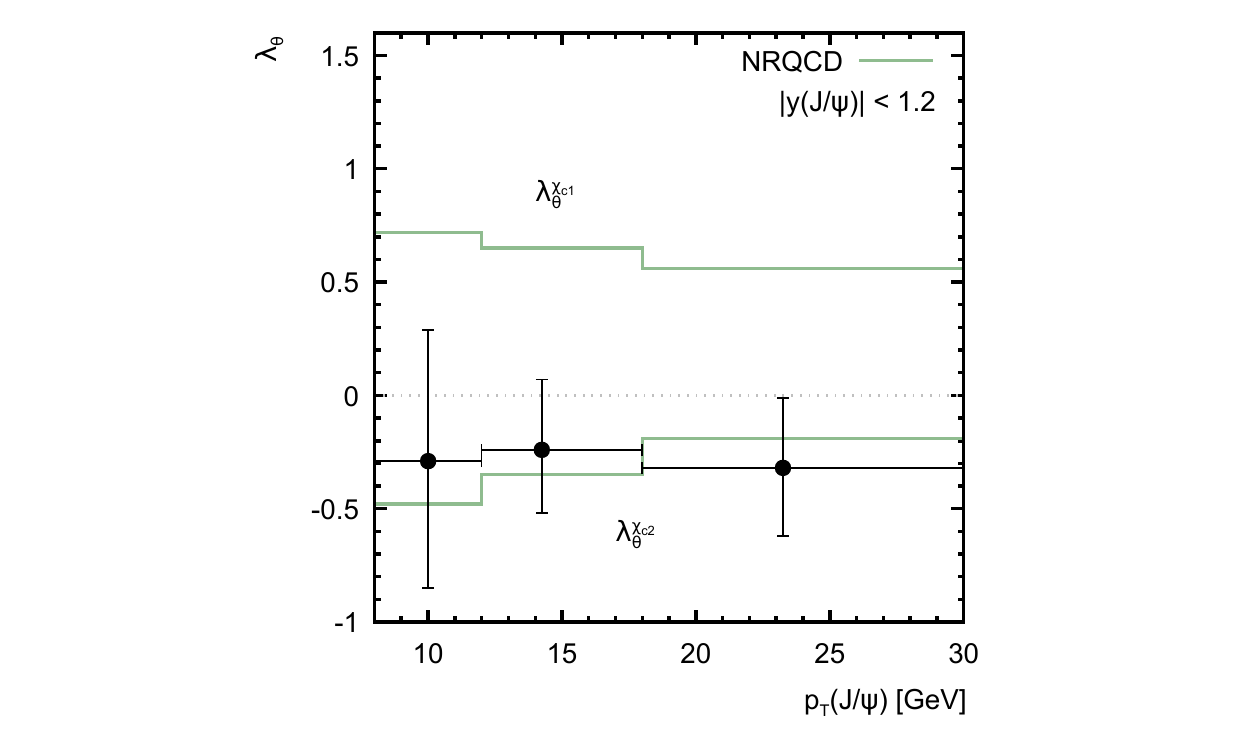} 
\caption{The $\lambda_{\theta}^{\chi_{c2}}$ values
determined according to correlations (5) --- (7) 
when the $\lambda_{\theta}^{\chi_{c1}}$ is fixed
to our predictions (left panel) or NRQCD ones (right panel).
The NRQCD predictions are taken from CMS paper\cite{1}.}
\label{fig5}
\end{center}
\end{figure}

\end{document}